\documentstyle[11pt,twoside,jltp]{article}

\input {epsf.sty}
\title{Analysis of Strong-Coupling Parameters for Superfluid $^{3}$He}

\author{H. Choi, J.P. Davis, J. Pollanen, T.M. Haard, W.P. Halperin\address{Department of Physics and Astronomy, Northwestern University, Evanston, IL 60208, USA}}

\runninghead{H. Choi {\em et al.}}{Analysis of Strong-Coupling Parameters for Superfluid $^{3}$He}
       
\begin{document}

\begin{abstract}
Superfluid $^{3}$He experiments show strong deviation from the weak-coupling limit of the Ginzburg-Landau theory, and this discrepancy grows with increasing pressure. Strong-coupling contributions to the quasiparticle interactions are known to account for this effect and they are manifest in the five $\beta$-coefficients of the fourth order Ginzburg-Landau free energy terms.  The Ginzburg-Landau free energy also has a coefficient $g_{z}$ to include magnetic field coupling to the order parameter.  From NMR susceptibility experiments, we find the deviation of $g_{z}$ from its weak-coupling value to be negligible at all pressures.  New results for the pressure dependence of four different combinations of $\beta$-coefficients, $\beta_{345}$, $\beta_{12}$, $\beta_{245}$, and $\beta_{5}$ are calculated and comparison is made with theory.

\end{abstract}

\maketitle

%Include this space if you do not use sections in your document.
%\vspace{0.3in}

\section{INTRODUCTION}
In the Ginzburg-Landau theory, the free energy can be expressed in even powers of the order parameter. A $p$-wave pairing superfluid, due to the complexity of its order parameter, has five fourth order terms, in addition to one second order term and a field coupling term in the presence of magnetic field.\cite{Thu87} The coefficients of these terms are manifest in thermodynamic properties of the superfluid such as transition temperature, specific heat, and magnetization. The coefficient of the second order term, $\alpha$, is determined solely from the transition temperature $T_{c}$, $\alpha=N(0)(1-T/T_{c})$ where $N(0)$ is the zero energy density of states of the Fermi liquid. The field coupling term, $g_{z}$, is related to the $^{3}$He-$B$ magnetization, $M_{B}$ as $g_{z} \propto M_{B}$. However, there are not five independent sets of measurements on superfluid $^{3}$He to unambiguously determine the coefficients of the five fourth order terms, the $\beta_{i}$'s. In principle, it is possible to obtain a fifth constraint from suitably precise measurements of the $AB$ surface tension at pressures less than the polycritical point\cite{Thu91}.

In the weak-coupling limit (wc), $g_{z}$ and the $\beta_{i}$'s are well-defined.\cite{Thu87} The weak-coupling value of $g_{z}$ is given by,
\begin{equation}
g_{z}^{\mathrm wc} = {7 \zeta(3) \over 48} {N(0) \over (\pi k_{B} T_{c})^{2}} \left( {\gamma \hbar \over 1+F_{0}^{a}} \right)^{2},
\end{equation} 
where $\gamma$ is the gyromagnetic ratio and $F_{0}^{a}$ is a Landau parameter in Fermi liquid theory. The $\beta_{i}$'s are given by -2$\beta_{1}^{\mathrm wc}$=$\beta_{2}^{\mathrm wc}$=$\beta_{3}^{\mathrm wc}$=$\beta_{4}^{\mathrm wc}$=-$\beta_{5}^{\mathrm wc}$=2$\beta_{0}$ where
\begin{equation}
\beta_{0} = {7 \zeta(3) \over 240} {N(0) \over (\pi k_{B} T_{c})^{2}}.
\end{equation} However, $^{3}$He behaves like a weak-coupling superfluid only in the limit of zero pressure and corrections to these coefficients due to strong-coupling are not fully understood.\cite{Sau81} In this work, we present magnetization measurements using NMR to extract $g_{z}$ up to 29 bar and calculate four known combinations of $\beta$ parameters from previous experiments including unpublished results from the NMR $g$-shift.

Historically, measuring $g_{z}$ has been a difficult task for two reasons. The first is that there is a discrepancy between the $^{3}$He-$B$ magnetization measurements using two different techniques\cite{Web77}, SQUID based static measurements\cite{sta} and NMR based dynamic measurements\cite{dyn}. The second is that the presence of the $A$-phase masks the magnetization of the $B$-phase above $T_{AB}$ and makes it impossible to directly measure $g_{z}$ at $T_{c}$. High precision measurements of the magnetization presented here allow us to extrapolate from below $T_{AB}$ up to $T_{c}$ and determine $g_{z}$.

\section{EXPERIMENT}
$^{3}$He magnetization is measured using a pulsed NMR technique in a 0.12 T magnetic field over the pressure range of 0-29 bar with a transverse field of 1.1 Oe and a pulse duration of 15 $\mu$s, resulting in a tipping angle of 20 degrees. The primary thermometer used in the experiment was a melting curve thermometer with Greywall temperature scale\cite{Gre86}. Pt susceptibility and solid $^{3}$He magnetization thermometers were also used to provide high-resolution temperature measurements at lower temperatures. A more detailed description of the experiment can be found elsewhere\cite{Haa01}.

\section{RESULTS AND DISCUSSION}

\begin{figure}[t]
%%%%%%%%%%%%%%%%%   F I G U R E  1   %%%%%%%%%%%%%%%%%%
\centerline{\epsfxsize0.75\hsize\epsffile{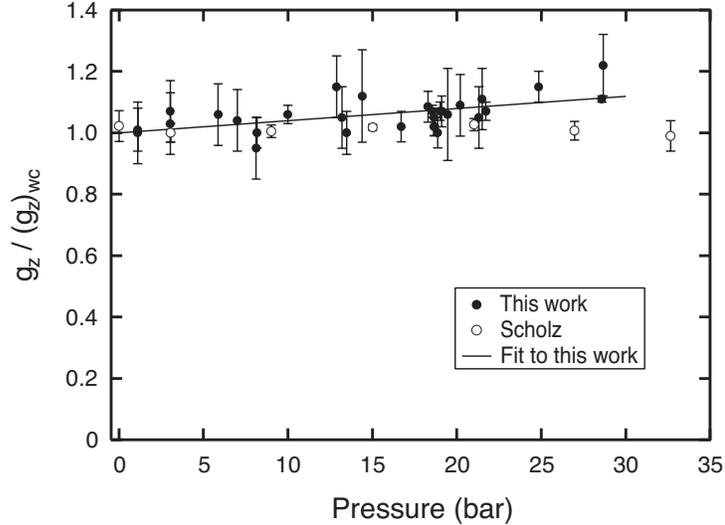}}
%%%%%%%%%%%%%%%%%%%%%%%%%%%%%%%%%%%%%%%%%%%%%
\begin{minipage}{1\hsize}
\caption{Pressure dependence of normalized Ginzburg-Landau parameter, $\hat g_{z}$, at $T_{c}$. Solid circles are our measurements in a 0.12 T magnetic field. Scholz {\em et al}.'s data are plotted in open circles for comparison. $g_{z}$ is equal to its weak-coupling value at all pressures within experimental uncertainty.}
\end{minipage}
\end{figure}
\noindent

As mentioned earlier, the presence of $A$-phase makes it impossible to directly measure the $B$-phase magnetization at $T_{c}$ and determine $g_{z}$. However, the magnetization and its field dependence has been calculated\cite{Fis86,Fis88,Moo93} and extrapolation of our measurements of the magnetization\cite{Haa01} up to $T_{c}$ is made in conjunction with these calculations. We define $\hat g_{z}$ as 
\begin{equation}
\hat g_{z} \equiv {g_{z} \over g_{z}^{\mathrm wc}}={{dm \over dt} \over ({dm \over dt})^{\mathrm wc}}{\beta_{B} \over \beta_{B}^{\mathrm wc}}
\end{equation}
where $m=M_{B}/M_{N}$, $t=T/T_{c}$ and $\beta_{B}=\beta_{12}+\beta_{345}/3$. In determining $\hat g_{z}$ from the experiment, it is helpful to know that $dm/dt$ is not field dependent in the Ginzburg-Landau limit. Combining this equation with the Ginzburg-Landau value of the magnetization at $T_{c}$\cite{Rai95}, we obtain $m(t)$ as a function of $\hat g_{z}$, $\beta_{345}/\hat g_{z}$, and $\beta_{B}$. We adopted the Mermin-Stare convention, that is, $\beta_{ij}=\beta_{i}+\beta_{j}$. $\beta_{345}/\hat g_{z}$ is obtained from the transverse NMR $g$-shift, $g$\cite{Haa01,Kyc94}, through the relation\cite{Gre76},
\begin{equation}
\frac{\beta_{345}}{\hat g_{z}}=\beta_{345}^{\mathrm wc} \frac{1}{(1+F_{0}^{a})^{2}} \left( \frac{C_{N}}{\Delta C_{B}}\right) \frac{\nu_{B \parallel}^{2}}{1-t} \left( \frac{\hbar}{2 \pi k_{B} T_{c}} \right)^{2} \frac{1}{g}
\label{Eq5}
\end{equation}
where $\Delta C_{B}/C_{N}$ is the $^{3}$He-$B$ specific heat jump\cite{Gre86}, and $\nu_{B \parallel}^{2}/(1-t)$ is the slope of the $B$-phase longitudinal resonance frequency\cite{Ran96}. $\beta_{B}$ comes directly from the $^{3}$He-$B$ specific heat jump. We can then solve for $\hat g_{z}$ at each temperature with the measured value of $m(T)$. $\hat g_{z}(T)$ varies smoothly near $T_{AB}$ and $\hat g_{z}$ can be extrapolated to $T_{c}$. For our analysis $F_{0}^{a}$ is taken from Ramm {\em et al}.\cite{Ram70}.

In Fig. 1, we show the final result of the analysis and plot $\hat g_{z}(T_{c})$ at various pressures. Data from Scholz {\em et al}.\cite{Sch81} are also analyzed in the same way and plotted for comparison. Our data for $\hat g_{z}$ show a weak pressure dependence, $\hat g_{z}$ increasing with pressure, whereas Scholz {\em et al}.'s\cite{Sch81} data shows no pressure dependence. In our case, the measurement is performed at a magnetic field of 0.12 T, higher than that of the Scholz {\em et al}. experiment, 0.06 T, resulting in lower $T_{AB}$'s. Therefore, a wider range for extrapolation was required in our case, especially at high pressures. This may be the source for the discrepancy between two measurements. In both cases, $\hat g_{z}$ is nearly unity at all pressures within the experimental uncertainty. This result is consistent with the conclusions reported by Hahn {\em et al}.\cite{Hah98} from static magnetization measurements.

%%%%%%%%%%%%%%%%%%%%% TABLE %%%%%%%%%%%%%%%%%%%%%
\begin{table}
{
\large
\centerline{
    \begin{tabular}{ccc||ccccccccc}
   &${P}\over{\mathrm(bar)}$& & &$\frac{\beta_{345}}{\beta_0}$ & & $ \frac{\beta_{12}}{\beta_0} $ & & $
    \frac{\beta_{245}}{\beta_0}
    $ & & $\frac{\beta_5}{\beta_0}$&\\*[-10pt]
    &&&&&&&&&&&\\ \hline\hline
    &&&&&&&&&&&\\*[-10pt]
    &wc & & & 2 & & 1& & 2& & -2&\\ \hline
    &&&&&&&&&&&\\*[-10pt]
    &0 & & & 2.11 && 0.92 && 1.90 && -1.84& \\
    %1 & 1.86 & 0.97 & 1.84 & -1.82& \\*[-3pt]
    %2 & 1.68 & 0.99 & 1.78 & -1.81 &\\*[-3pt]
    &3 &&& 1.56 && 1.01 && 1.74 && -1.81& \\
    %4 && 1.47 && 1.01 && 1.70 && -1.81& \\*[-4pt]
    %5 && 1.41 && 1.01 && 1.66 && -1.81& \\*[-3pt]
    &6 &&& 1.36 && 1.01 && 1.63 && -1.82& \\
    %7 && 1.32 && 1.00 && 1.60 && -1.82& \\*[-3pt]
    %8 && 1.29 && 1.00 && 1.57 && -1.83& \\*[-3pt]
    &9 &&& 1.26 && 0.99 && 1.55 && -1.84& \\
    %10 && 1.24 && 0.98 && 1.52 && -1.85& \\*[-3pt]
    %11 && 1.23 && 0.98 && 1.50 && -1.86& \\*[-3pt]
    &12 &&& 1.21 && 0.97 && 1.48 && -1.87& \\
    %13 && 1.20 && 0.96 && 1.46 && -1.88 &\\*[-3pt]
    %14 && 1.19 && 0.96 && 1.44 && -1.89 &\\*[-3pt]
    &15 &&& 1.18 && 0.95 && 1.41 && -1.89& \\
    %16 && 1.17 && 0.95 && 1.39 && -1.90 &\\*[-3pt]
    %17 && 1.15 && 0.94 && 1.37 && -1.90 &\\*[-3pt]
    &18 &&& 1.14 && 0.94 && 1.35 && -1.91& \\
    %19 && 1.13 && 0.93 && 1.34 && -1.92 &\\*[-3pt]
    %20 && 1.12 && 0.93 && 1.32 && -1.93 &\\*[-3pt]
    &21 &&& 1.10 && 0.93 && 1.30 && -1.93& \\
    %22 && 1.09 && 0.92 && 1.28 && -1.94 &\\*[-3pt]
    %23 && 1.08 && 0.92 && 1.27 && -1.95 &\\*[-3pt]
    &24 &&& 1.06 && 0.92 && 1.25 && -1.96 &\\
    %25 && 1.05 && 0.92 && 1.24 && -1.97 &\\*[-3pt]
    %26 && 1.03 && 0.91 && 1.23 && -1.97 &\\*[-3pt]
    &27 &&& 1.02 && 0.91 && 1.21 && -1.98 &\\
    %28 && 1.00 && 0.91 && 1.20 && -1.99 &\\*[-3pt]
    %29 && 0.99 && 0.91 && 1.19 && -2.00 &\\*[-3pt]
    &30 &&& 0.97 && 0.91 && 1.18 && -2.01 &\\
    %31 &&  && &  1.16 && -2.02 &\\*[-3pt]
    %32 &&  && &  1.15 && -2.02 &\\*[-3pt]
    &33 &&& 0.90 && 0.90 &&  1.14 && -2.03 &\\
    %34 &  & &  1.12 & -2.03\\
    \end{tabular}
}
\begin{minipage}{1\hsize}
\vspace{10pt}
Table 1. Four combinations of the $\beta_{i}$'s measured from the NMR $g$-shift\cite{Kyc94}, specific heat jump\cite{Gre86}, quadratic suppression of the $B$-phase by magnetic field\cite{Tan91}, and $A_{1}$-$A_{2}$ splitting\cite{Isr84}. $\hat g_{z}=1$ was used in calculating $\beta_{345}$ and subsequently $\beta_{12}$.
\end{minipage}
}\end{table}

Once $\hat g_{z}$ is determined, four combinations of the $\beta_{i}$'s can be determined. $\beta_{345}$ can be calculated from Eq.(5) with $\hat g_{z}$, $\Delta C_{B}/C_{N}$, $g$, and $\nu_{B \parallel}^{2}/(1-t)$ known. $\Delta C_{B}/C_{N}$ is a direct measure of $\beta_{B}$ and with $\beta_{345}$, $\beta_{12}$ can be calculated subsequently.

$\beta_{245}$ and $\beta_{5}$ are determined independent of the measurement of $g_{z}$. Below the polycritical pressure, the quadratic suppression of the $B$-phase by a magnetic field\cite{Tan91}, $g(\beta)$ is used along with the asymmetry ratio of the $A_{1}$-$A_{2}$ splitting\cite{Isr84} which gives the ratio $\beta_{5}/\beta_{245}$. Above the polycritical pressure, $\beta_{245}$ is calculated directly from $A$-phase specific heat jump, $\Delta C_{A}/C_{N}$\cite{Gre86}. These results are tabulated in Table 1.

\begin{figure}[t]
%%%%%%%%%%%%%%%%%   F I G U R E  2   %%%%%%%%%%%%%%%%%%
\centerline{\epsfxsize0.75\hsize\epsffile{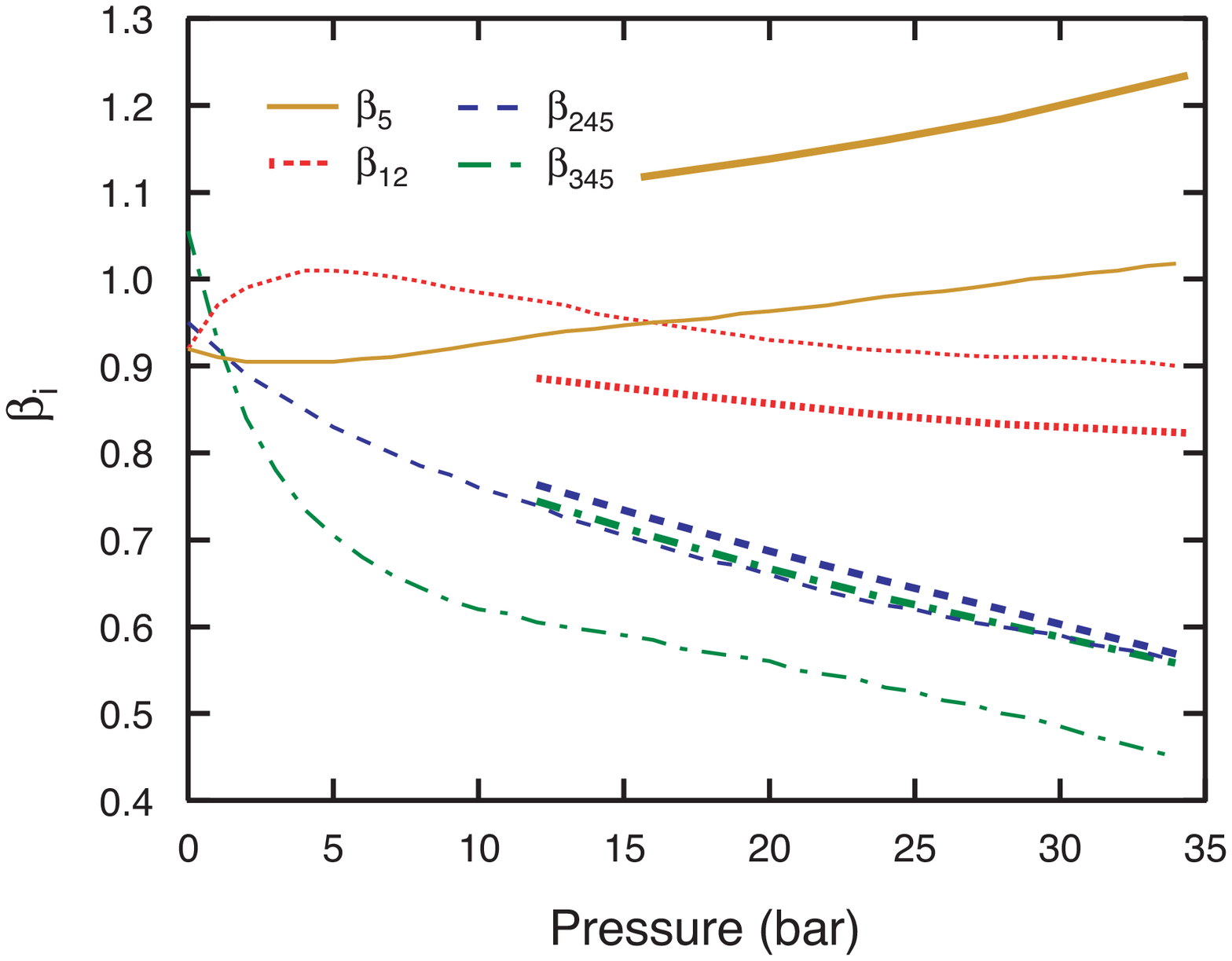}}
%%%%%%%%%%%%%%%%%%%%%%%%%%%%%%%%%%%%%%%%%%%%%
\begin{minipage}{1\hsize}
\vspace{6pt}
Fig. 2. Four combinations of the $\beta_{i}$'s from Table 1, normalized to their weak-coupling values, are compared with calculation by Sauls and Serene\cite{Sau81}. The set of thin lines that spans the entire pressure range is from experiment and the set of bold lines above 12 bar is from the theory.
\end{minipage}
\end{figure}
\noindent

At low pressures, the four combinations of the $\beta_{i}$'s recover their weak-coupling values to within a few percent. As the pressure grows, however, the deviation from this weak-coupling limit increases due to strong-coupling of the quasiparticle interactions. Sauls and Serene\cite{Sau81} calculated the strong-coupling corrections to the $\beta_{i}$'s above 12 bar and from their calculation we obtain theoretical values for $\beta_{12}$, $\beta_{245}$, $\beta_{345}$, and $\beta_{5}$. We plot their calculation and the measurements of these four combinations of the $\beta_{i}$'s normalized to their weak-coupling values in Fig. 2. The absolute values of the theory and experiment are only qualitatively consistent. However, the pressure dependence is in very good agreement above the pressure of 12 bar.

In summary, we have measured the magnetization of $^{3}$He and obtained $g_{z}$ 
at various pressures in the Ginzburg-Landau limit. We find that $g_{z}$ is equal to its weak-coupling value at all pressures within experimental error. Knowledge of $g_{z}$ is essential to the interpretation of the NMR $g$-shift in terms of the $\beta_{i}$'s of the Ginzburg-Landau theory, more specifically $\beta_{345}$. From previous experiments, three other combinations of $\beta$ parameters can then be calculated. These four $\beta$ parameter combinations are in qualitative agreement with the theory of Sauls and Serene\cite{Sau81} and the pressure dependence appears to be in excellent agreement. This observation will be helpful in resolving the ambiguity associated with determining all five $\beta$ parameters independently\cite{Cho06}.

\section{ACKNOWLEDGMENTS}
This research is supported by National Science Foundation, DMR-0244099. We thank D. Rainer and J.A. Sauls for their helpful discussion.

\end{document}